\documentclass[]{emulateapj}

\begin{document}

\title{New chemical profiles for the asteroseismology of ZZ Ceti stars}

\author{L. G. Althaus$^{1,2}$,
        A. H. C\'orsico$^{1,2}$,
        A. Bischoff-Kim$^{3}$,
        A. D. Romero$^{1,2}$,
        I. Renedo$^{4,5}$,  
        E. Garc\'{\i}a--Berro$^{4,5}$,
        and M. M. Miller Bertolami$^{1,2}$}

\affil{$^1$Facultad de Ciencias Astron\'omicas y Geof\'{\i}sicas, 
           Universidad Nacional de La Plata, 
           Paseo del Bosque s/n, 
           (1900) La Plata, 
           Argentina\\
       $^2$Instituto de Astrof\'{\i}sica de La Plata, 
           IALP (CCT La Plata), 
           CONICET-UNLP\\
       $^3$Department of Chemistry, Physics and Astronomy, CBX 82, 
           Georgia College \& State University, Milledgeville, GA 31061, 
           USA\\
       $^4$Departament de F\'\i sica Aplicada,
           Universitat Polit\`ecnica de Catalunya,
           c/Esteve Terrades 5, 
           08860 Castelldefels,
           Spain\\
       $^5$Institute for Space Studies of Catalonia, 
           c/Gran Capit\`a 2--4, 
           Edif. Nexus 104, 
           08034 Barcelona,
           Spain\\}
\email{althaus@fcaglp.unlp.edu.ar}

\begin{abstract}
We compute  new chemical profiles for  the core and  envelope of white
dwarfs  appropriate for pulsational  studies of  ZZ Ceti  stars. These
profiles  are  extracted from  the  complete  evolution of  progenitor
stars,  evolved through  the main  sequence and  the thermally-pulsing
asymptotic giant branch (AGB)  stages, and from time-dependent element
diffusion during  white dwarf evolution. We discuss  the importance of
the initial-final mass relationship  for the white dwarf carbon-oxygen
composition.  In particular, we find that the central oxygen abundance
may be underestimated by about 15\% if the white dwarf mass is assumed
to be the hydrogen-free core  mass before the first thermal pulse.  We
also discuss the importance for  the chemical profiles expected in the
outermost  layers  of  ZZ  Ceti   stars  of  the  computation  of  the
thermally-pulsing  AGB  phase  and  of  the  phase  in  which  element
diffusion is relevant. We find  a strong dependence of the outer layer
chemical stratification  on the stellar  mass.  In particular,  in the
less massive models, the  double-layered structure in the helium layer
built  up during  the thermally-pulsing  AGB phase  is not  removed by
diffusion  by the  time the  ZZ Ceti  stage is  reached.   Finally, we
perform adiabatic pulsation  calculations and discuss the implications
of our new chemical profiles for the pulsational properties of ZZ Ceti
stars.   We find  that  the  whole $g-$mode  period  spectrum and  the
mode-trapping properties  of these  pulsating white dwarfs  as derived
from our new chemical  profiles are substantially different from those
based on chemical profiles widely used in existing asteroseismological
studies. Thus,  we expect the asteroseismological  models derived from
our chemical  profiles to be significantly different  from those found
thus far.
\end{abstract}

%_____________________________________________________________________

\section{Introduction}  
\label{intro}  

Pulsating DA (H-rich atmospheres) white dwarfs, also called ZZ Ceti or
DAV stars, are  the most numerous class of  degenerate pulsators, with
over  143 members  known  today  (Winget \&  Kepler  2008). Since  the
discovery of  the first ZZ  Ceti star, HL  Tau 76, by  Landolt (1968),
there  has been a  continuous effort  to model  the interior  of these
variable stars.  ZZ Ceti stars are found within a very narrow strip of
effective  temperatures ($10\,500$  K $\lesssim  T_{\rm  eff} \lesssim
12\,500$  K).   They  are  characterized by  multiperiodic  brightness
variations  of  up to  $0.30$  mag  caused  by spheroidal,  non-radial
$g$-modes of low  degree ($\ell \leq 2$) with  periods between 100 and
1200 s.  The  driving mechanism thought to excite  the pulsations near
the  blue  edge  of  the  instability  strip  is  the  $\kappa-\gamma$
mechanism  that takes place  in the  hydrogen partial  ionization zone
(Dolez \&  Vauclair 1981; Dziembowski  \& Koester 1981; Winget  et al.
1982).  Also,  the ``convective driving'' mechanism  has been proposed
--- first by Brickhill (1991) and later re-examined by Goldreich \& Wu
(1999). It appears to be the  responsible of mode driving once a thick
convection zone has developed at the stellar surface.

The comparison of  the observed pulsation periods in  white dwarfs and
the periods  computed for appropriate theoretical  models (white dwarf
asteroseismology) allows  to infer  details of their  origin, internal
structure and  evolution (Winget \& Kepler 2008;  Fontaine \& Brassard
2008).  In  particular, the stellar  mass, the thickness of  the outer
envelopes, the core chemical composition, magnetic fields and rotation
rates can be  determined from the observed periods.   In addition, the
asteroseismology  of ZZ  Ceti stars  is a  valuable tool  for studying
axions (Isern  et al.  1992;  C\'orsico et al.  2001;  Bischoff-Kim et
al.  2008; Isern et al.   2010) and crystallization (Montgomery et al.
1999; C\'orsico et  al.  2004, 2005; Metcalfe et  al.  2004; Kanaan et
al.  2005). Finally, the temporal  changes in the observed periods can
help  detect  planets  orbiting   around  white  dwarfs  (Mullally  et
al. 2008).

The first published complete set of DA white dwarf models suitable for
asteroseismology  was  that  of  Tassoul  et  al.   (1990).   A  large
parameter space  was explored  in such a  monumental study, and  for a
long time  (since the early  eighties) this set of  models represented
the state-of-the-art  in the area.  The pulsation properties  of these
models were  thoroughly explored  in a series  of important  papers by
Brassard et  al.  (1991, 1992a,  1992b). As important as  these models
were at  that time,  they suffer from  a number of  shortcomings.  For
instance, the  core of  the models  is made of  pure  carbon,  while stellar
evolution calculations indicate that cores of typical white dwarfs are
made  of a  mixture of   carbon and oxygen.  Also,  the carbon/helium (C/He)  
and helium/hydrogen (He/H) chemical
interfaces are modeled on the basis of the assumption of the diffusive
equilibrium in  the ``trace element approximation'',  an approach that
involves  a  quasi-discontinuity  in  the  chemical  profiles  at  the
transition regions  which, in  turn, leads to  peaked features  in the
Brunt-V\"ais\"al\"a  frequency and  exaggerated  mode-trapping effects
(C\'orsico  et al.   2002a, 2002b).   These models  were  employed for
asteroseismological inferences of the DAVs G 226$-$29 (Fontaine et al.
1992) and GD 154 (Pfeiffer et  al.  1996).  More recently, Pech et al.
(2006) and Pech \&  Vauclair (2006) have presented asteroseismological
analysis  on HL  Tau 76  and  G 185$-$32,  respectively, by  employing
similar DA white dwarf models, although with updated input physics.

The models of Bradley  (1996) constituted a substantial improvement in
the  field.    These  models  have  carbon-oxygen   cores  in  varying
proportions,  and  the C/He  and  He/H  chemical  interfaces are  more
realistic. Perhaps the most severe  shortcoming of these models is the
(unrealistic)  ramp-like  shape  of  the core  carbon-oxygen  chemical
profiles.   These DA  models  were  the basis  of  the very  important
asteroseismological studies  on the DAVs G 29-38  (Bradley \& Kleinman
1997), G  117$-$B15A and  R 548  (Bradley 1998), GD  165 and  L 19$-$2
Bradley (2001), and G 185$-$32 (Bradley 2006).

The next step in improving the modeling of DAVs was given by C\'orsico
et al. (2002b) and Benvenuto et al. (2002a), who employed evolutionary
models  characterized by  He/H  chemical interfaces  resulting from  a
time-dependent  element  diffusion  treatment  (Althaus  \&  Benvenuto
2000), and  the carbon-oxygen  core chemical structure  extracted from
the evolutionary computations  of Salaris et al.  (1997).   The use of
very  smooth   outer  chemical  interfaces,  as   shaped  by  chemical
diffusion, revealed  that the use  of the trace  element approximation
turns  out to  be inappropriate  to model  the shape  of  the chemical
interfaces in a  DA white dwarf.  This grid of  models was employed in
an asteroseismological study of  G 117$-$B15A (Benvenuto et al. 2002a).
For these  sequences, the starting configurations for  the white dwarf
evolution were  obtained through an  artificial procedure, and  not as
result of evolutionary computations of the progenitor stars.

Recently,  Castanheira \&  Kepler  (2008, 2009)  have  carried out  an
extensive  asteroseismological study  of  DAVs by  employing DA  white
dwarf models similar to those of Bradley (1996), but with a simplified
treatment  of the  core  chemical structure,  by somewhat  arbitrarily
fixing  the central  abundances to  50 \%  oxygen and  50 \%  carbon.   The He/H
chemical interfaces adopted for  these models are a parametrization of
the realistic chemical  profiles resulting from time-dependent element
diffusion (Althaus et al.  2003). The study includes the ``classical''
DAVs and also the recently discovered  SDSS DAVs. In total, 83 ZZ Ceti
stars are analyzed.  An important  result of these studies is that the
thickness of the H envelopes  inferred from asteroseismology is in the
range $10^{-4}  \gtrsim M_{\rm H}/M_*  \gtrsim 10^{-10}$, with  a mean
value of  $M_{\rm H}/M_*=  5 \times 10^{-7}$.   This suggests  that an
important  fraction of  DAs characterized  by  envelopes substantially
thinner than  predicted by the standard evolution  theory could exist,
with  the consequent important  implications for  the theory  of white
dwarf formation.   However, these results  are preliminary and  do not
include the  possible effects  of realistic carbon-oxygen  profiles on
the asteroseismological fits.

Almost simultaneously  with the study of Castanheira  \& Kepler (2008,
2009), Bischoff-Kim et al.  (2008) performed a new asteroseismological
study on  G 117$-$B15A and  R 548 by  employing DA white  dwarf models
similar to those  employed by Castanheira \& Kepler  (2008, 2009), but
incorporating realistic core chemical profiles according to Salaris et
al. (1997).  The results of this work are in reasonable agreement with
previous  studies on  these  ZZ  Ceti stars.   However,  the mass  and
effective  temperatures found  by Bischoff-Kim  et al.   (2008)  for G
117$-$B15A  are  rather  high   (especially  the  mass,  at  $0.66  \,
M_{\sun}$).  Recently, Bogn\'ar et  al.  (2009) have employed the same
asteroseismological modeling  to study the  pulsations of the  ZZ Ceti
star  KUV  02464+3239.   Finally,  Bischoff-Kim (2009)  presented  the
results  of an  asteroseismological  analysis of  two  DAVs with  rich
pulsation spectrum, G 38$-$29 and  R 808 based on similar models, with
parametrized, smooth  ramp-like core profiles.  These  models are able
to reproduce the observed  period spectra reasonably well, though some
assumptions  about the  $m$ and  $\ell$ identification  of  modes were
made.

White  dwarf  stellar  models  with realistic  chemical  profiles  are
crucial  to  correctly  assess   the  adiabatic  period  spectrum  and
mode-trapping properties of the DAVs,  which lies at the core of white
dwarf  asteroseismology   (Brassard  et  al.    1992a;  Bradley  1996;
C\'orsico et  al.  2002a).   In this paper,  we compute for  the first
time  consistent chemical  profiles for  both the  core {\sl  and} the
envelope of  white dwarfs with various stellar  masses appropriate for
detailed asteroseismological  fits of  ZZ Ceti stars.   These chemical
profiles  are  extracted  from  the  full and  complete  evolution  of
progenitor   stars  from   the  zero   age  main   sequence,   to  the
thermally-pulsing and mass-loss phases  on the asymptotic giant branch
(AGB),  and from time-dependent  element diffusion  predictions during
the white dwarf stage\footnote{The  chemical profiles for the core and
envelope of our models appropriate  for ZZ Ceti stars are available at
{\tt  http://www.fcaglp.unlp.edu.ar/evolgroup}}.  These  profiles will
be  valuable in  conducting future  asteroseismological studies  of ZZ
Ceti  stars  that  intend   to  include  realistic  chemical  profiles
throughout the interior of white dwarfs. To assess the impact of these
new  profiles on  the  theoretical pulsational  inferences, we  perfom
adiabatic  pulsation computations  and compare  the  resulting periods
with the pulsational inferences based on the most widely used chemical
profiles in existing asteroseismological fits.

The  paper  is organized  as  follows.   In  Sect.  \ref{physics},  we
provide a description of the input physics assumed in the evolutionary
calculations   of  relevance   for  the   chemical   composition.   In
Sect. \ref{mimf}  we discuss the importance of  the initial-final mass
relationship for  the expected white  dwarf carbon-oxygen composition.
The resulting chemical profiles are described at some length in Sect.
\ref{profiles}.  The implications of our new chemical profiles for the
pulsational  properties  of  ZZ  Ceti  stars are  discussed  in  Sect.
\ref{pulsation}. We conclude in Sect. \ref{conclusions} by summarizing 
our findings.  

%_____________________________________________________________________

\section{Input physics} 
\label{physics} 

The chemical profiles presented in this paper have been extracted from
full and complete evolutionary  calculations for both the white dwarfs
and their  progenitor stars, by using  an updated version  of the {\tt
LPCODE} stellar evolutionary  code --- see Althaus et  al.  (2005) and
references  therein.  This code  has recently  been employed  to study
different  aspects of  the evolution  of low-mass  stars, such  as the
formation  and evolution  of DA white dwarfs (Renedo et al. 2010), H-deficient white  dwarfs,  PG 1159  and
extreme  horizontal  branch  stars   (Althaus  et  al.   2005;  Miller
Bertolami \& Althaus  2006; Miller Bertolami et al.   2008; Althaus et
al.  2009a),  as well  as the evolution  of He-core white  dwarfs with
high  metallicity progenitors (Althaus  et al.   2009b).  It  has also
been  used to  study  the initial-final-mass  relation  in Salaris  et
al. (2009),  where a  test and comparison  of {\tt LPCODE}  with other
evolutionary codes has also been made.  Details of {\tt LPCODE} can be
found in  these works. In what  follows, we comment on  the main input
physics that are relevant  for his  work.  We  assume the metallicity  of progenitor stars  to be
$Z=0.01$.

\begin{table}
\centering
\caption{Initial  and  final  stellar mass  (in solar units), and  the 
         central  oxygen  abundance (mass  fraction)  left after  core
         helium  burning, and after  Rayleigh-Taylor rehomogenization.
         The progenitor metallicity is $Z=0.01$.}
\begin{tabular}{cccc}
\hline
\hline
$M_{\rm ZAMS}$ & $M_{\rm WD}$ &  $X_{\rm O}$(CHB) & $X_{\rm O}$(RT)\\
\hline
1.00 & 0.5249 & 0.702 & 0.788 \\
1.50 & 0.5701 & 0.680 & 0.686 \\
1.75 & 0.5932 & 0.699 & 0.704 \\
2.00 & 0.6096 & 0.716 & 0.723 \\
2.25 & 0.6323 & 0.747 & 0.755 \\
2.50 & 0.6598 & 0.722 & 0.730 \\
3.00 & 0.7051 & 0.658 & 0.661 \\
3.50 & 0.7670 & 0.649 & 0.655 \\
4.00 & 0.8373 & 0.635 & 0.641 \\
5.00 & 0.8779 & 0.615 & 0.620 \\
\hline
\hline
\end{tabular}
\label{tableini}
\end{table}  

The {\tt LPCODE} evolutionary  code considers a simultaneous treatment
of non-instantaneous mixing (and  extra-mixing if present) and burning
of  elements (Althaus  et al.   2003).  The  nuclear  network accounts
explicitly  for the  following elements:  $^{1}$H,  $^{2}$H, $^{3}$He,
$^{4}$He, $^{7}$Li, $^{7}$Be,  $^{12}$C, $^{13}$C, $^{14}$N, $^{15}$N,
$^{16}$O,  $^{17}$O,  $^{18}$O,  $^{19}$F,  $^{20}$Ne  and  $^{22}$Ne,
together with  34 thermonuclear reaction  rates of the  pp-chains, CNO
bi-cycle, helium burning,  and carbon  ignition that are  identical to those
described in Althaus et al.   (2005), with the exception of $^{12}$C$\
+\ $p$ \rightarrow \ ^{13}$N +  $\gamma \rightarrow \ ^{13}$C + e$^+ +
\nu_{\rm  e}$ and  $^{13}$C(p,$\gamma)^{14}$N, which  were  taken from
Angulo et  al.  (1999).  The  $^{12}$C($\alpha,\gamma)^{16}$O reaction
rate, of special relevance for the carbon-oxygen stratification of the
emerging  white dwarf,  was taken  from  Angulo et  al.  (1999).   The
standard  mixing  length  theory  for  convection ---  with  the  free
parameter $\alpha=1.61$  --- has been  adopted.  With this  value, the
present luminosity and effective temperature  of the Sun, at an age of
4\,570  Myr,  are  reproduced  by  {\tt LPCODE}  when  $Z=0.0164$  and
$X=0.714$  are  adopted ---  in  agreement  with  the $Z/X$  value  of
Grevesse \& Sauval (1998).

Except   for   the    evolutionary   stages   corresponding   to   the
thermally-pulsing  AGB phase,  we  have considered  the occurrence  of
extra-mixing  episodes beyond each  convective boundary  following the
prescription of Herwig  et al.  (1997).  We treated  extra-mixing as a
diffusion   process   by  assuming   that   mixing  velocities   decay
exponentially  beyond  each  convective  boundary.   Specifically,  we
assume  a  diffusion coefficient  given  by  $D_{\rm  OV}= D_{\rm  O}\
\exp(-2z/f H_{\rm P})$, where $H_{\rm P}$ is the pressure scale height
at the  convective boundary, $D_{\rm O}$ is  the diffusion coefficient
of unstable regions  close to the convective boundary,  and $z$ is the
geometric distance from the edge of the convective boundary (Herwig et
al.   1997).  We  adopted  $f=0.016$  in all  our  sequences, a  value
inferred from the width of the upper main sequence.  The occurrence of
extra-mixing  episodes during core  helium burning  largely determines
the final chemical composition of the white dwarf core. In this sense,
our treatment of time-dependent  extra-mixing episodes predicts a core
chemical stratification similar to that predicted by the phenomenon of
``semiconvection''  during  central  helium burning,  which  naturally
yields the  growth of  the convective core  (Straniero et  al.  2003).
Finally, the breathing pulse  instability occurring towards the end of
core helium  burning was suppressed  --- see Straniero et  al.  (2003)
for a discussion  of this point. In our  simulations, breathing pulses
have been  suppressed by gradually  decreasing the parameter  $f$ from
the moment the  helium convective core starts to  receed (which occurs
once  the  helium  abundance  in  the core  decreases  below  $\approx
0.05-0.1$).  At this  stage, the  gradual suppression  of extra-mixing
toward the  end of core helium  burning bears no  consequences for the
final chemical stratification.

Other  physical  ingredients  considered   in  {\tt  LPCODE}  are  the
radiative opacities  from the OPAL project (Iglesias  \& Rogers 1996),
including carbon- and oxygen-rich  composition, supplemented at low temperatures
with the molecular opacities  of Alexander \& Ferguson (1994).  During
the white dwarf regime, the metal mass fraction $Z$ in the envelope is
not  assumed  to be  fixed.   Instead,  it  is specified  consistently
acording to the prediction of element diffusion.  To account for this,
we have considered radiative  opacities tables from OPAL for arbitrary
metallicities.  For effective temperatures  less than 10,000 K we have
included the  effects of molecular opacitiy by  assuming pure hydrogen
composition from  the computations of Marigo \&  Aringer (2009).  This
assumption  is  justified  because  element diffusion  leads  to  pure
hydrogen envelopes in cool white dwarfs.  The conductive opacities are
those of Cassisi  et al.  (2007), and the  neutrino emission rates are
taken from Itoh et al.  (1996)  and Haft et al.  (1994).  For the high
density  regime characteristics  of  white dwarfs,  we  have used  the
equation of state of Segretain  et al.  (1994), which accounts for all
the important contributions  for both the liquid and  solid phases ---
see Althaus et al.  (2007) and references therein.

In this study, we have considered the distinct physical processes that
are  responsible for  changes in  the chemical  abundance distribution
during  white  dwarf  evolution.   In  particular,  element  diffusion
strongly  modifies the chemical  composition profile  throughout their
outer  layers.  As  a  result of  diffusion  processes, our  sequences
developed pure  hydrogen envelopes,  the thickness of  which gradually
increases  as evolution  proceeds.  We  have  considered gravitational
settling  as  well as  thermal  and  chemical  diffusion ---  but  not
radiative levitation, which is relevant only for the hottest and 
brightest post-AGB and early white-dwarf cooling stages for determining 
the surface  composition --- of $^1$H, $^3$He, $^4$He,
$^{12}$C, $^{13}$C, $^{14}$N and  $^{16}$O, see Althaus et al.  (2003)
for details.   Our treatment of  time-dependent diffusion is  based on
the multicomponent gas treatment presented in Burgers (1969).  In {\tt
LPCODE}, diffusion becomes operative once the wind limit is reached at
high  effective temperatures  (Unglaub  \& Bues  2000).  In  addition,
abundance changes  resulting from residual nuclear  burning --- mostly
during the  hot stages  of white dwarf  evolution --- have  been taken
into account in our  simulations.  Finally, we considered the chemical
rehomogenization  of  the  inner  carbon- oxygen  profile  induced  by
Rayleigh-Taylor (RT)  instabilities following Salaris  et al.  (1997).
These  instabilities arise  because of  the positive  molecular weight
gradients  that  remain  above  the  flat  chemical  profile  left  by
convection during core helium burning.

%_____________________________________________________________________

\section{The importance of the initial-final mass relationship}
\label{mimf}

As mentioned,  chemical profiles appropriate for DA  white dwarfs have
been  derived from  the full  evolutionary calculations  of progenitor
stars for solar  metallicity.  To this end, the  complete evolution of
ten evolutionary sequences with initial stellar mass ranging from 1 to
$5\,   M_{\sun}$   were   computed   from   the   ZAMS   through   the
thermally-pulsing and mass-loss  phases on the AGB and  finally to the
domain  of planetary  nebulae.  In  Table \ref{tableini}  we  list the
stellar masses of the resulting white dwarfs, together with the inital
masses  of the progenitor  stars on  the ZAMS.   Also listed  in Table
\ref{tableini} is the central oxygen abundance both at the end of core
He  burning  and after  chemical  rehomogenization by  Rayleigh-Taylor
instabilities.

\begin{figure} 
\centering 
\includegraphics[clip,width=250pt]{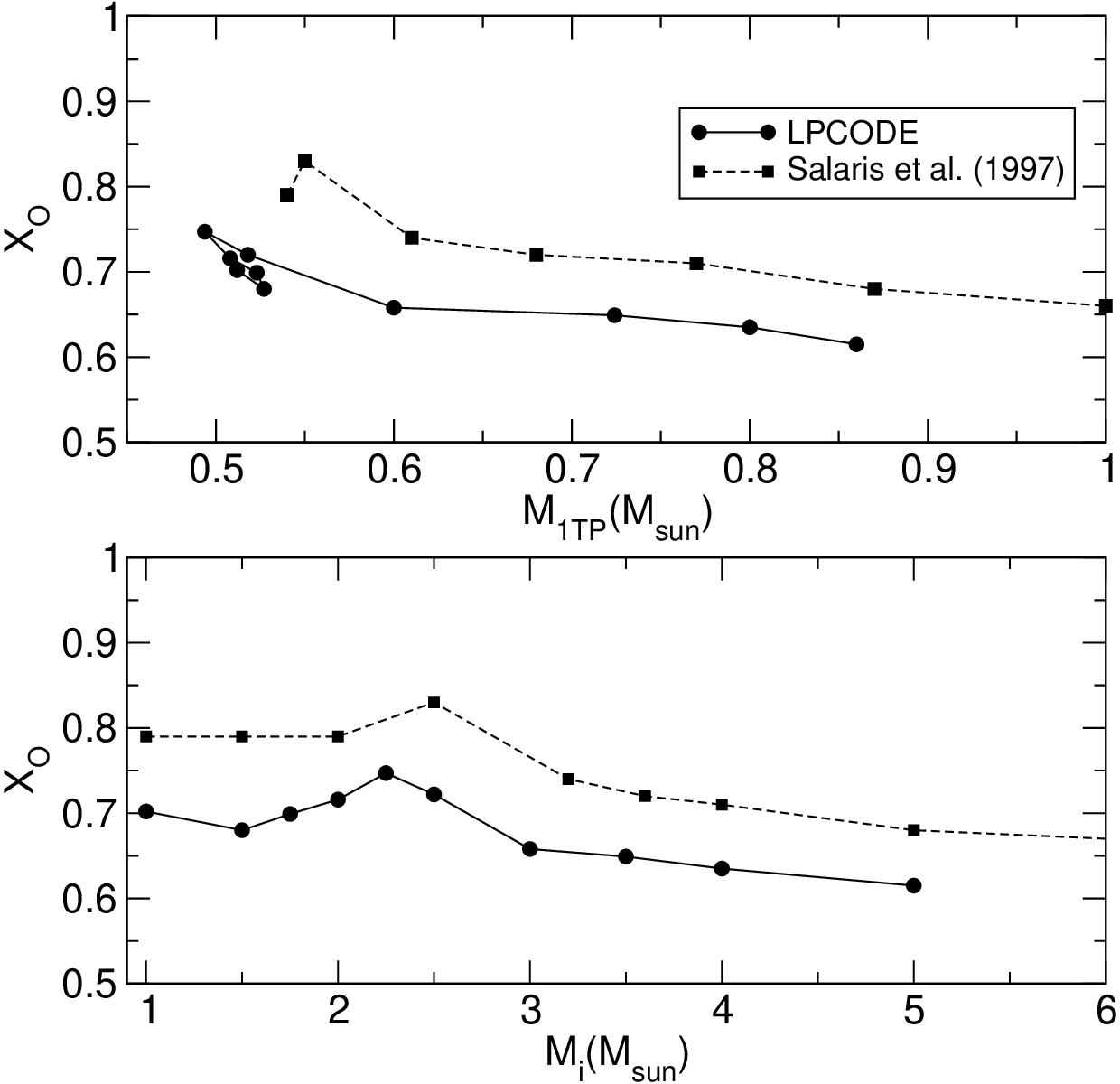} 
\caption{Central  oxygen  abundance  (mass  fraction)  left after core 
         helium burning  in terms of both the  hydrogen-free core mass
         right before  the first thermal  pulse (upper panel)  and the
         initial stellar  mass on the  ZAMS (lower panel).   The solid
         lines show our  results while the dashed lines  show those of
         Salaris et al.  (1997)}
\label{oxiini} 
\end{figure} 

We  mention that  extra-mixing  episodes were  disregarded during  the
thermally-pulsing  AGB phase.   In particular,  a strong  reduction (a
value of $f$ much smaller  than 0.016) of extra-mixing episodes at the
base of  the pulse-driven convection zone is  supported by simulations
of the $s-$process abundance patterns  (Lugaro et al.  2003) and, more
recently,  by  observational  inferences  of  the  initial-final  mass
relation (Salaris et al.  2009). As  a result, it is expected that the
mass of  the hydrogen-free  core of our  sequences gradually  grows as
evolution proceeds through the thermally-pulsing AGB.  This is because
a strong reduction  or suppression of extra-mixing at  the base of the
pulse-driven convection zone strongly inhibits the occurrence of third
dredge-up,  thus favoring the  growth of  the hydrogen-free  core.  We
considered mass-loss episodes during the core helium burning stage and
on the red giant branch  following Schr\"oder \& Cuntz (2005).  During
the AGB and thermally-pulsing AGB  phases, we adopted the maximum mass
loss rate between  the prescription of Schr\"oder \&  Cuntz (2005) and
that of Vassiliadis \& Wood (1993).  In the case of a strong reduction
of third  dredge-up, as occurred in  our sequences, mass  loss plays a
major role in determining the  final mass of the hydrogen-free core at
the  end of  the TP-AGB  evolution,  and thus  the initial-final  mass
relation (Weiss \& Ferguson 2009).

\begin{figure} 
\centering 
\includegraphics[clip,width=250pt]{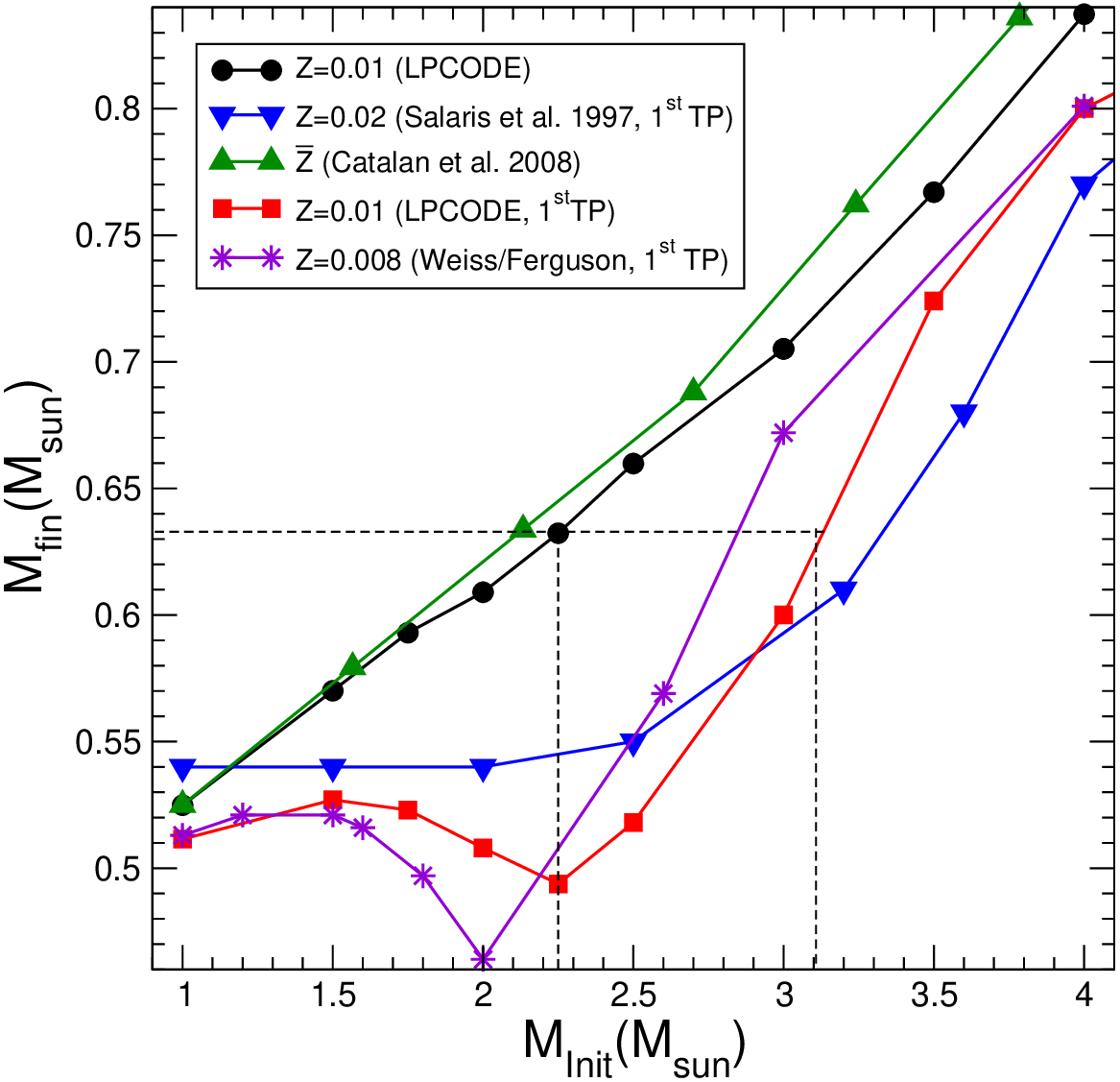} 
\caption{Initial-final  mass relationships: The  final mass  given  by 
         the  hydrogen-free core  mass  is depicted  in  terms of  the
         initial  mass  of  the  progenitor star.  In  addition to the 
         observational   data  from   open   clusters  (Catal\'an   et
         al.  2008),  upwards   triangles,  we  show  the  theoretical
         predictions  given by  our  calculations at  the  end of  the
         TP-AGB phase and before  the first thermal pulse (circles and
         squares,  respectively).  Also shown  are the  predictions of
         Weiss \& Ferguson (2009) and Salaris et al. (1997) before the
         first   thermal  pulse   (stars   and  downwards   triangles,
         respectively).}
\label{mimfinal} 
\end{figure}

We  begin by examining  Fig. \ref{oxiini}  which displays  the central
oxygen abundance by mass fraction left after core helium burning.  The
upper panel shows  the predicted central oxygen abundance  in terms of
the  hydrogen-free core  mass right  before the  first  thermal pulse,
while  the lower panel  shows this  quantity in  terms of  the initial
stellar mass  on the  ZAMS.  The predictions  of our  calculations ---
solid lines --- are compared with  those of Salaris et al.  (1997) ---
dashed lines.  Note the qualitatively good agreement between both sets
of   calculations.    We   recall   that   the   final   carbon-oxygen
stratification  of  the  emerging  white  dwarf depends  on  both  the
efficiency  of the  $^{12}$C($\alpha,\gamma)^{16}$O reaction  rate and
the occurrence of extra-mixing episodes  toward the late stage of core
helium  burning.   In  particular,  the  systematically  lower  oxygen
abundances  of our  models are  due  mostly to  our use  of the  cross
section for the $^{12}$C($\alpha,\gamma)^{16}$O reaction rate given by
Angulo et al. (1997), which is smaller than the rate of Caughlan et al
(1985) adopted  by Salaris  et al.   (1997).  Note  that both  sets of
calculations predict a  maximum in the central oxygen  abundance at an
initial mass of $M \approx 2.5 M_{\sun}$.

\begin{figure} 
\centering 
\includegraphics[clip,width=250pt]{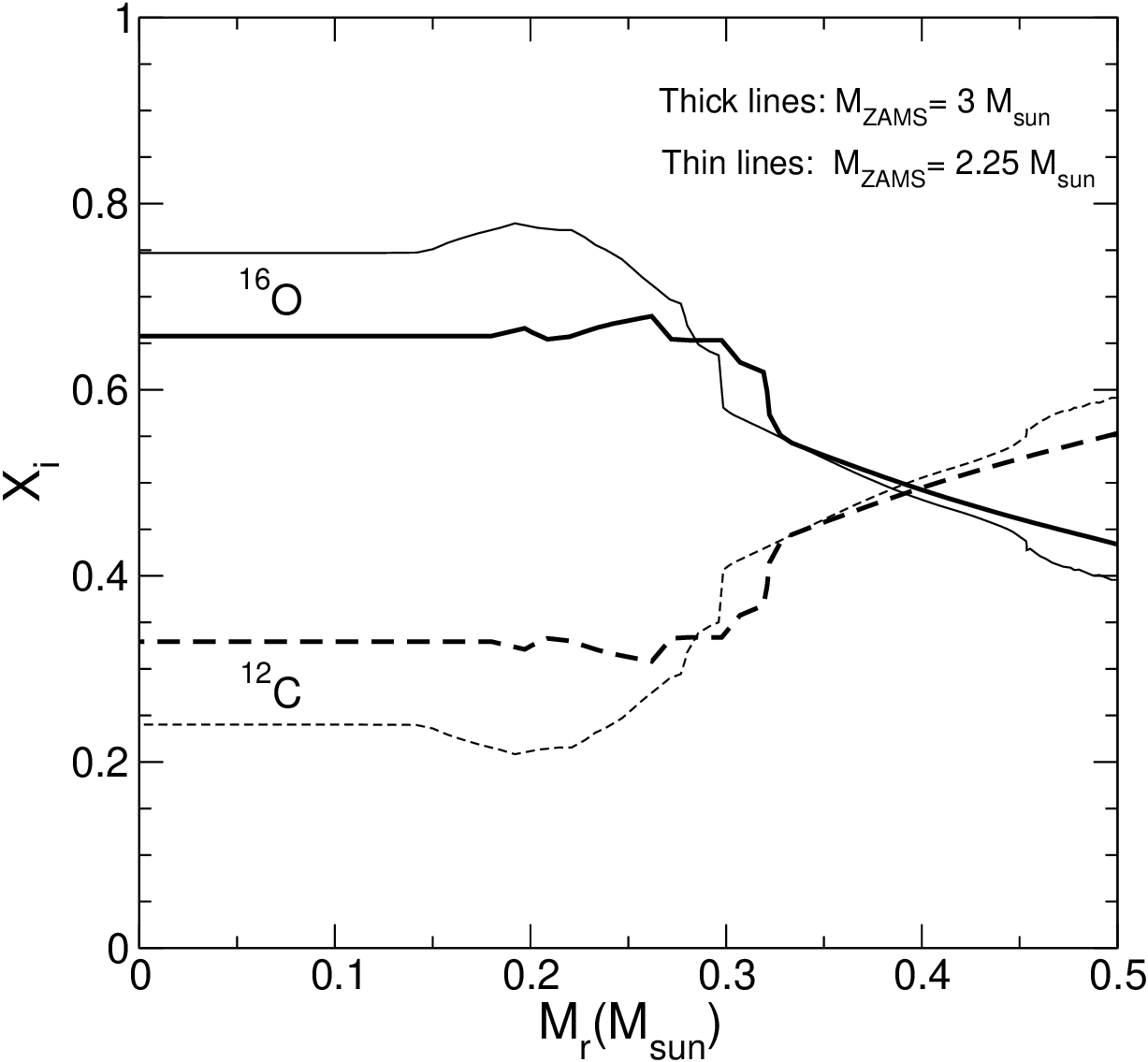} 
\caption{Inner carbon  and  oxygen abundance  by mass for the  $0.63\, 
         M_{\sun}$   white   dwarf   resulting  from   two   different
         progenitors  that lead  to  the same  white  dwarf mass.  The
         chemical profiles correspond to progenitor stars with initial
         stellar mass  of 3  and  $2.25 \,  M_{\sun}$ (thick  and thin
         lines,   respectively),  and   at   stages  before   chemical
         rehomogenization by Rayleigh-Taylor instability.}
\label{perfilcomparo} 
\end{figure} 

A  careful   computation  of   the  evolutionary  stages   during  the
thermally-pulsing   AGB   and   the   resulting   initial-final   mass
relationship is an important aspect concerning  the final  carbon-oxygen  composition of  the
white   dwarf   core.   This   can   be   seen   by  inspecting   Fig.
\ref{mimfinal},   where   various   theoretical   initial-final   mass
relationships, giving the  mass of the hydrogen-free core  in terms of
the ZAMS  mass of the progenitor,  are plotted.  The  results shown in
this  figure   include  the  predictions  of   our  full  evolutionary
calculations at the end of  the thermally-pulsing AGB phase and at the
beginning   of  the   first  thermal   pulse  (circles   and  squares,
respectively).  Our  relationships are compared with  those of Salaris
et al.  (1997) and Weiss \& Fergusson (2009), both at the beginning of
the   first  thermal   pulse.    We  also   show  the   semi-empirical
initial-final mass relationship of  Catal\'an et al.  (2008), based on
white dwarfs in open clusters  and in common proper motions pairs with
metallicities close  to $Z= 0.01$,  the metallicity we assume  for the
progenitor stars of  our sequences.  In view of  the discussion of the
preceeding  paragraph,   note  the  increase   in  the  mass   of  the
hydrogen-free  core  during the  thermally-pulsing  AGB  stage.  As  a
result,  the  initial-final  mass  relationship  by  the  end  of  the
thermally-pulsing AGB becomes  markedly different from that determined
by the mass of the  hydrogen-free core before the first thermal pulse.
For the carbon-oxygen  composition expected in a white  dwarf, this is
an important issue.  Indeed, if  the mass of the hydrogen-free core is
assumed to be essentially the mass of the resulting white dwarf, it is
clear that a white dwarf with a given mass may correspond to different
progenitor stellar masses depending  on the assumed initial-final mass
relationship.   For instance,  from  our theoretical  initial-to-final
mass  relationships,  a white  dwarf  with  $0.63  \, M_{\sun}$  would
correspond  to a  progenitor  star with  a  stellar mass  of $2.25  \,
M_{\sun}$  if the initial-final  mass relationship  is assessed  at an
advanced stage in the thermally-pulsing  AGB phase, or $3 \, M_{\sun}$
if the white  dwarf mass is assumed to  be the hydrogen-deficient core
mass right  before the  first thermal pulse.   In particular,  we note
that Salaris et al. (1997) adopt the mass of the hydrogen-free core at
the start  of the  first thermal  pulse as the  mass of  the resulting
white dwarf.

\begin{figure} 
\centering 
\includegraphics[clip,width=250pt]{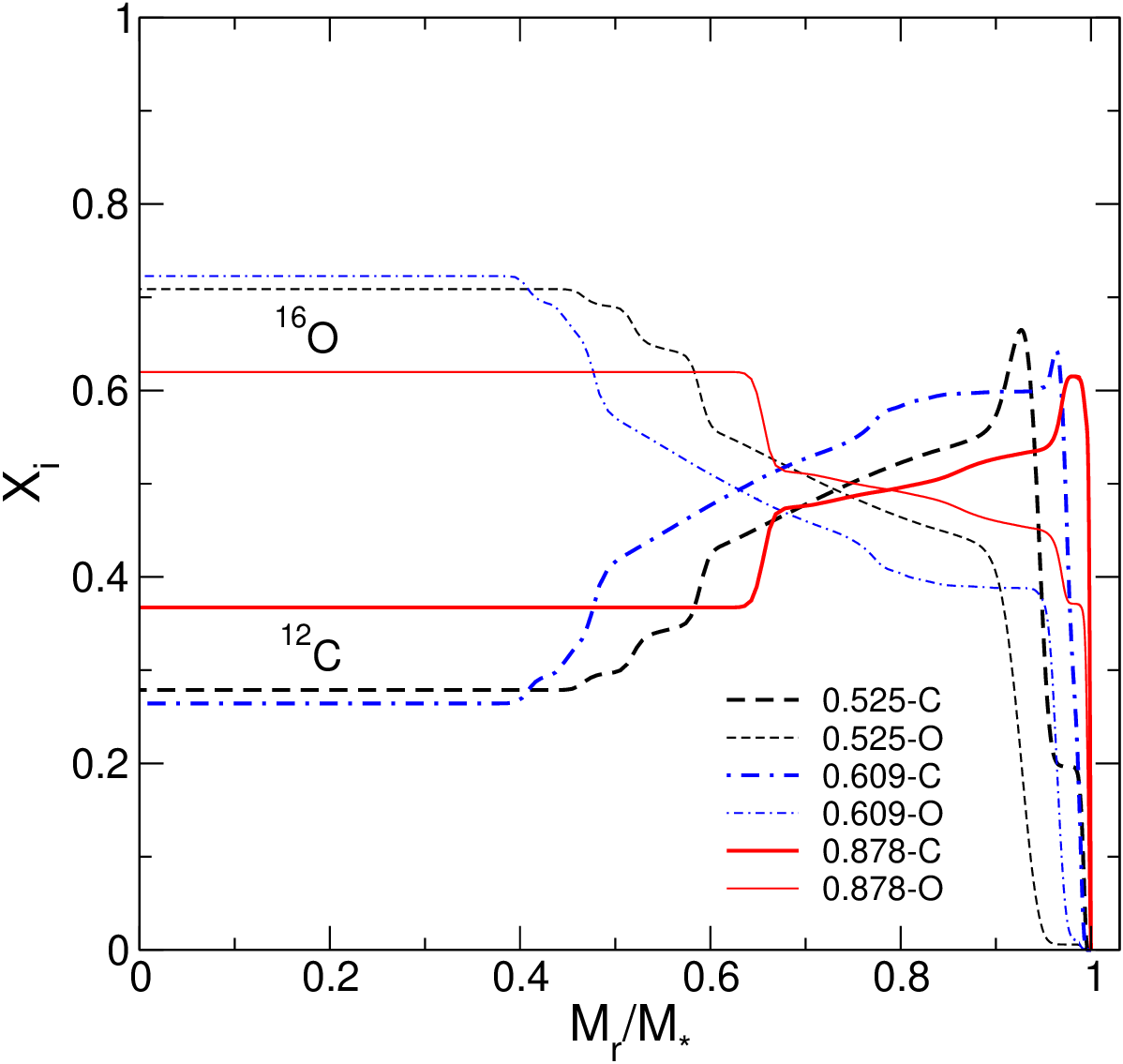} 
\caption{Inner  carbon  and  oxygen  abundance  by mass for the 0.525, 
         0.609, and $0.878 \,  M_{\sun}$ white dwarf models at $T_{\rm
         eff} \approx 12,000K$, and  after  chemical  rehomogenization  
         by Rayleigh-Taylor instabilities.}
\label{perfilesco} 
\end{figure} 

The implication of  this dichotomy in the mass  of the progenitor star
for the  white dwarf carbon-oxygen composition is  illustrated in Fig.
\ref{perfilcomparo}.  We  show the  inner carbon and  oxygen abundance
distribution for the $0.63 \, M_{\sun}$ white dwarf resulting from the
two  different  progenitors  discussed previously.   For  illustrative
purposes,  chemical rehomogenization by  Rayleigh-Taylor instabilities
has not been  considered in this particular example.   The thick lines
display  the chemical  profile  for  the progenitor  star  with $3  \,
M_{\sun}$  characterized   before  the   first  thermal  pulse   by  a
hydrogen-free core  of $0.63  \, M_{\sun}$.  The  thin lines  show the
chemical profile for the $2.25  \, M_{\sun}$ progenitor which leads to
the same  white dwarf mass  but after evolution has  proceeded through
the  thermally-pulsing AGB phase  (see Fig.\ref{mimfinal}).   Note the
different chemical  profiles expected  in both cases.   In particular,
the  central oxygen  abundance  may be  underestimated  by about  15\%
should the  white dwarf mass is  assumed to be  the hydrogen-free core
mass right before the first thermal pulse. Note that, however, this 
variation is an upper limit, and it would be less for other white-dwarf 
masses. Clearly, the initial-final
mass  relationship is  an  aspect that  has  to be  considered in  the
problem of the carbon-oxygen composition expected in a white dwarf, as
well as  in attempts at  constraining, from pulsational  inferences of
variable white dwarfs, the mixing  processes and the efficiency of the
$^{12}$C($\alpha,\gamma)^{16}$O  reaction rate in  the core  of helium
burning stars. Also,  it is important to realize  that a larger oxygen
abundance increases the cooling rate of the white dwarf because of the
lower  heat capacity and  because an  initial larger  oxygen abundance
reduces  the energy  release  by phase  separation on  cyrstallization
(Isern et al. 2000).

%_____________________________________________________________________

\section{The internal chemical profiles}
\label{profiles}

The carbon-oxygen stratification for  some selected models is shown in
Figs.   \ref{perfilcomparo} and  \ref{perfilesco}.  The  shape  of the
chemical    profiles    before    rehomogenization   is    given    in
Fig.  \ref{perfilcomparo}. Easily recognizable  are the  flat chemical
profiles in the  inner part of the core left  by convection during the
core  helium  burning,  the  bumps  resulting from  the  inclusion  of
extra-mixing  episodes  beyond  the  fully convective  core,  and  the
signatures of the outward-moving helium-burning shell after the end of
core helium burning.  Because of the larger temperatures in the helium
burning shell, the oxygen abundance  decreases in the outer regions of
the carbon-oxygen core.

\begin{figure} 
\centering 
\includegraphics[clip,width=250pt]{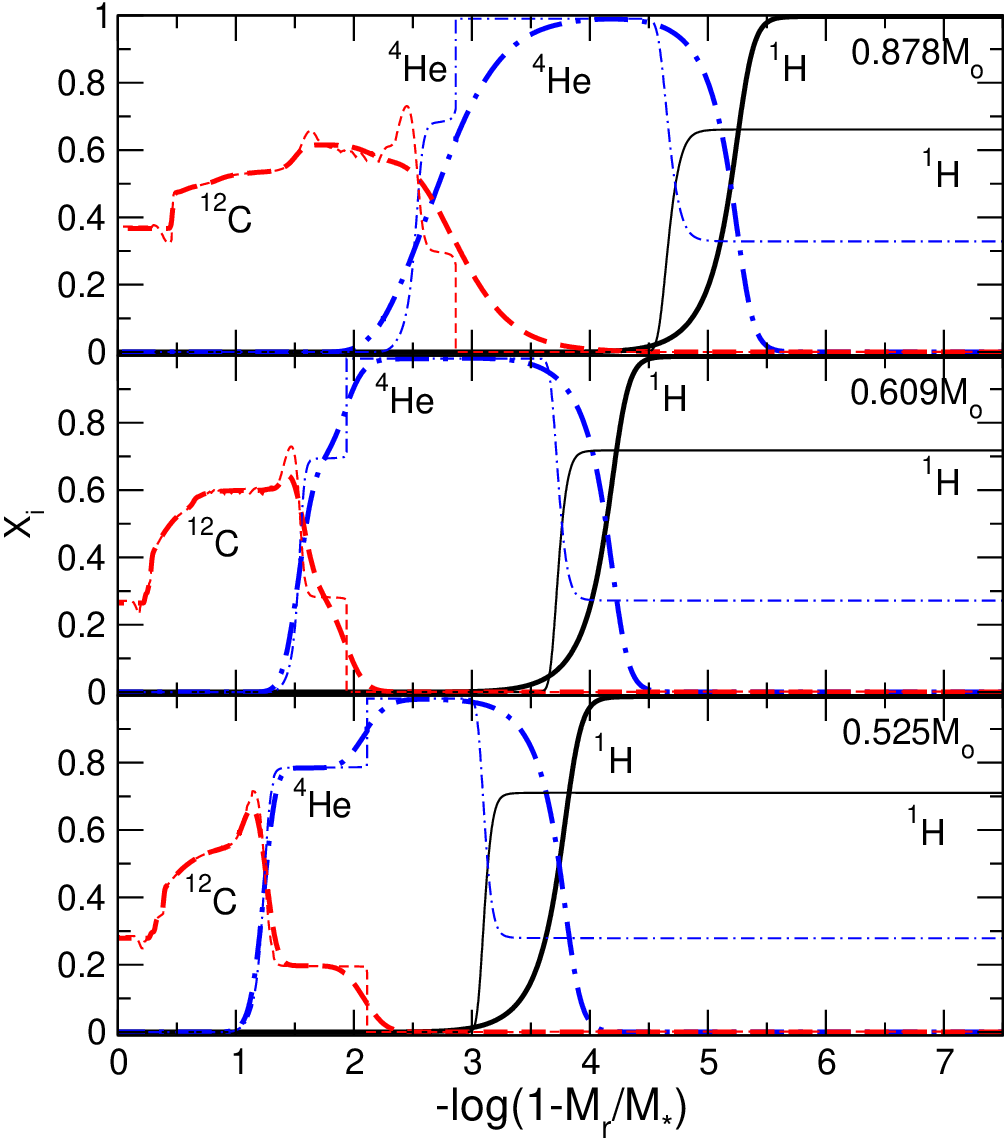} 
\caption{Abundance distribution of hydrogen, helium and carbon in terms 
         of the outer mass fraction for the 0.878, 0.609 and $0.525 \,
         M_{\sun}$  white dwarf  models  at two  selected stages  just
         after the  maximum effective  temperature point and  near the
         beginning  of  the ZZ  Ceti  regime  (thin  and thick  lines,
         respectively). }
\label{perfilouter} 
\end{figure} 

The expected chemical profiles of  some of our white dwarf models when
evolution  has  proceeded to  the  domain of  the  ZZ  Ceti stars  are
displayed  in   Fig.   \ref{perfilesco}.   At   this  stage,  chemical
rehomogenization by Rayleigh-Taylor  instabilities has already smeared
out the  bumps in the inner  profiles, leading to  quite extended flat
chemical  profiles. Note  the  dependence of  both  the core  chemical
abundances and the location of the chemical transitions on the stellar
mass. Pulsation  periods in white  dwarf models are very  sensitive to
the shapes and locations of  the chemical transions zones (see section
5).   This  emphasizes  the  need  for a  detailed  knowledge  of  the
progenitor history for a  realistic treatment of white dwarf evolution
and  pulsations.   The chemical  profiles  in  the outermost  regions,
resulting   from  prior   mixing   and  burning   events  during   the
thermally-pulsing AGB  phase, are  markedly modified by  the diffusion
processes  acting during  white dwarf  evolution, particularly  in the
case   of  more   massive   models,  where   chemical  diffusion   and
gravitational settling are notably more efficient.  This can be better
appreciated  in Fig. \ref{perfilouter},  where the  chemical abundance
distribution  for white dwarf  models of  different stellar  masses is
depicted in terms of the outer mass fraction.  These plots put special
emphasis in  the outer regions of  the model.  In this  figure and for
each stellar mass, thin lines show the chemical abundance distribution
at early  stages of white dwarf  evolution when diffusion  has not had
time   to  act.    The  signature   of  the   evolution   through  the
thermally-pulsing AGB stage on  the chemical profile, particularly the
formation  of the  helium-rich  buffer and  the underlying  intershell
region  rich in  helium  and carbon  ---  built up  during the  mixing
episode at  the last  AGB thermal pulse  --- are easily  visible.  The
presence of carbon in the intershell region stems from the short-lived
convective mixing  that has driven the carbon-rich  zone upward during
the peak of the last helium  pulse on the AGB.  Thick lines depict the
situation at advanced stages, near the ZZ Ceti instability strip, when
element  diffusion  has   strongly  modified  the  chemical  abundance
distribution and  resulted in the  formation of a thick  pure hydrogen
envelope plus  an extended  inner tail.  In  the more  massive models,
chemical  diffusion leads  to a  significant amount  of carbon  in the
helium  buffer   zone.   Also  near-discontinuities   in  the  initial
abundance distribution are smoothed out considerably by diffusion.

An important fact to note in Fig.  \ref{perfilouter} is the dependence
on  the  stellar  mass  of  the outer  layer  chemical  stratification
expected  in ZZ  Ceti  stars.  Indeed, for  the  more massive  models,
diffusion has  strongly modified the  chemical abundance distribution,
eroding the  intershell region by  the time evolution has  reached the
domain of the ZZ Ceti instability strip.  This is in contrast with the
situation encountered in our less massive models, where the intershell
region is not removed by  diffusion. This is because element diffusion
is  less  efficient  in  less  massive models  and  also  because  the
intershell   is  thicker,   with  the   subsequent   longer  diffusion
timescales.   Regarding white  dwarf asteroseismology,  these  are not
minor issues, since the presence  of a double-layered structure in the
helium-rich  layers is  expected  to affect  the theoretical  $g-$mode
period  spectrum  of  ZZ Ceti  stars,  as  it  does for  pulsating  DB
(He-rich)  white  dwarf stars  (Metcalfe  et  al.  2003). However,  we
mention  that the  thickness  of the  intershell  region also  depends
somewhat on the number of thermal pulses during the AGB experienced by
the progenitor star.

Finally,  we have  explored whether  the  shape of  the He/H  chemical
interface has a dependence on the thickness of the H envelope. This is
an important  issue, because predictions of  the exact value  of the H
envelope  mass are tied  to the  precise mass  loss history  along the
previous AGB and post-AGB phase, and particularly to the occurrence of
late  thermal   pulses.  We  have   generated  additional,  artificial
sequences   with  H  envelopes   much  thinner   than  those   of  our
models. These sequences  were created at high luminosities  from a hot
model  with $M_*=  0.609 M_{\odot}$.   This artificial  procedure took
place at luminosities high enough  as to ensure that the models become
physically sound at stages far before the domain of the ZZ Ceti stars.
We found that diffusion rapidly  leads to pure hydrogen envelopes, but
the shape  of the He/H chemical  interfaces by the  time evolution has
proceeded to  the ZZ Ceti stage,  is almost the  same independently of
the thickness of the H envelope.

%_____________________________________________________________________

\section{Pulsation properties: comparison with previous calculations}
\label{pulsation}

White dwarf asteroseismology is sensitive  to the precise shape of the
internal chemical  profiles.  The entire $g-$mode  period spectrum and
mode-trapping properties of pulsating  white dwarfs are very sensitive
to the fine details of the  chemical profiles of both the core and the
envelope of the star. This extreme sensitivity has been exploited with
some success in  several pulsation studies to infer  the core chemical
structure --- e.g., Metcalfe (2003) for the case of pulsating DB white
dwarfs  --- and  the thickness  of the  He and  H envelopes  --- e.g.,
Bradley (1998, 2001) for DAV stars.

\begin{figure}  
\centering  
\includegraphics[clip,width=230pt]{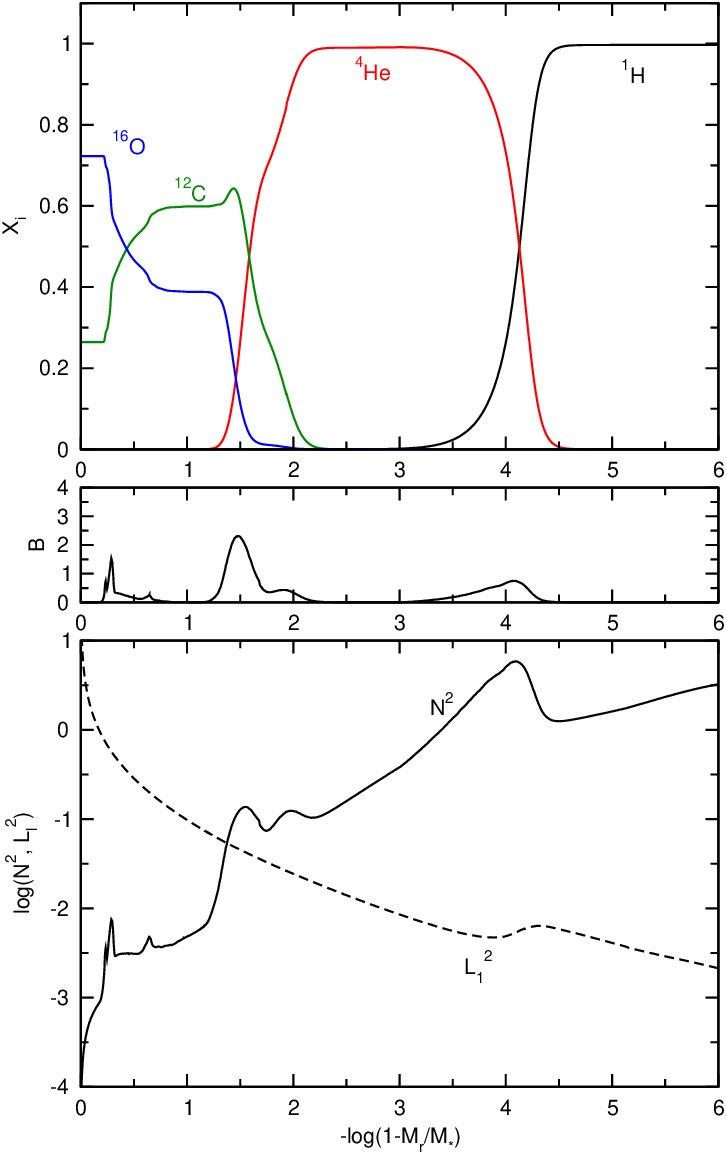}  
\caption{The internal chemical profile (upper panel), the  Ledoux term 
         $B$  (middle  panel),  and   the  logarithm  of  the  squared
         Brunt-V\"ais\"al\"a  ($N$) and Lamb  ($L_{\ell}$) frequencies
         (lower panel) in terms of the outer mass fraction ($-\log q$,
         where  $q \equiv  1-M_r/M_*$)  for dipole  modes ($\ell=  1$)
         corresponding to  a DA white dwarf model  with $M_*= 0.6096\,
         M_{\sun}$ and $T_{\rm eff} \sim 12\,000$ K.}
\label{x_b_n_LPCODE}  
\end{figure}  

\begin{figure}  
\centering  
\includegraphics[clip,width=230pt]{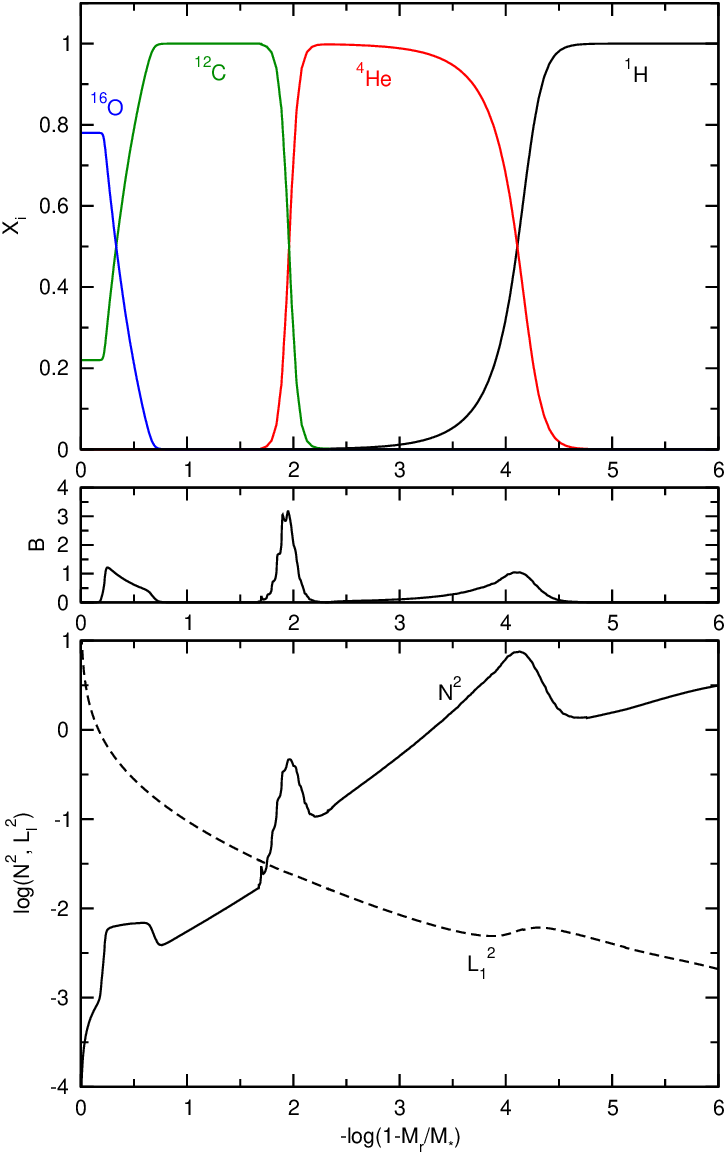}  
\caption{Same  as  in  Fig.  \ref{x_b_n_LPCODE}, but for  the case of 
         ramp-like core chemical profiles.}
\label{x_b_n_ramp-like}  
\end{figure}  
        
In  this  section  we  perform  a  comparison  between  the  pulsation
properties derived from  our new chemical profiles and  those based on
the most widely used chemical profiles in existing asteroseismological
studies.  To  assess the adiabatic  pulsation properties of  our white
dwarf models  we employ the  numerical code described in  C\'orsico \&
Althaus (2006).  We refer the  reader to that paper for details.  With
the  aim of simplifying  our analysis,  we elect  a template  DA white
dwarf model with $M_*=  0.6096\, M_{\sun}$, $T_{\rm eff} \sim 12\,000$
K, and a thick hydrogen  envelope ($M_{\rm H} \sim 10^{-4} M_*$). This
is a canonical model of a DAV star with an average mass located in the
middle  of   the  observed  ZZ   Ceti  instability  strip.    In  Fig.
\ref{x_b_n_LPCODE}  we  depict the  internal  chemical profile  (upper
panel),  the Ledoux  term (middle  panel),  and the  logarithm of  the
squared  Brunt-V\"ais\"al\"a and  Lamb frequencies  (lower  panel) for
dipole modes corresponding to our template model.  The Ledoux term and
the  Brunt-V\"ais\"al\"a frequency  are  computed as  in C\'orsico  \&
Althaus  (2006).   Our  model   is  characterized  by  three  chemical
transition regions: a double chemical  interface of oxygen and carbon located at
the  core region  ($0.2  \lesssim  -\log q  \lesssim  0.8$), a  triple
chemical interface  of oxygen, carbon, and  helium located at $1.2  \lesssim -\log q
\lesssim  2.3$,  and finally,  a  double  chemical  interface of  He/H
located at  $3.0 \lesssim  -\log q \lesssim  4.5$.  The  core chemical
profile is typical of situations in which extra mixing episodes beyond
the  fully convective  core  (like overshooting)  during  the core  He
burning  phase are  allowed to  operate.  The  smoothness of  the He/H
chemical  interface,  on  the  other   hand,  is  the  result  of  the
time-dependent  element diffusion processes.   The existence  of these
three chemical interfaces induces the  ``bumps'' in the profile of the
Brunt-V\"ais\"al\"a frequency.  The number  of these bumps, as well as
their heights and  widths, strongly affect the whole  structure of the
pulsation spectrum of the star.

\begin{figure}  
\centering  
\includegraphics[clip,width=230pt]{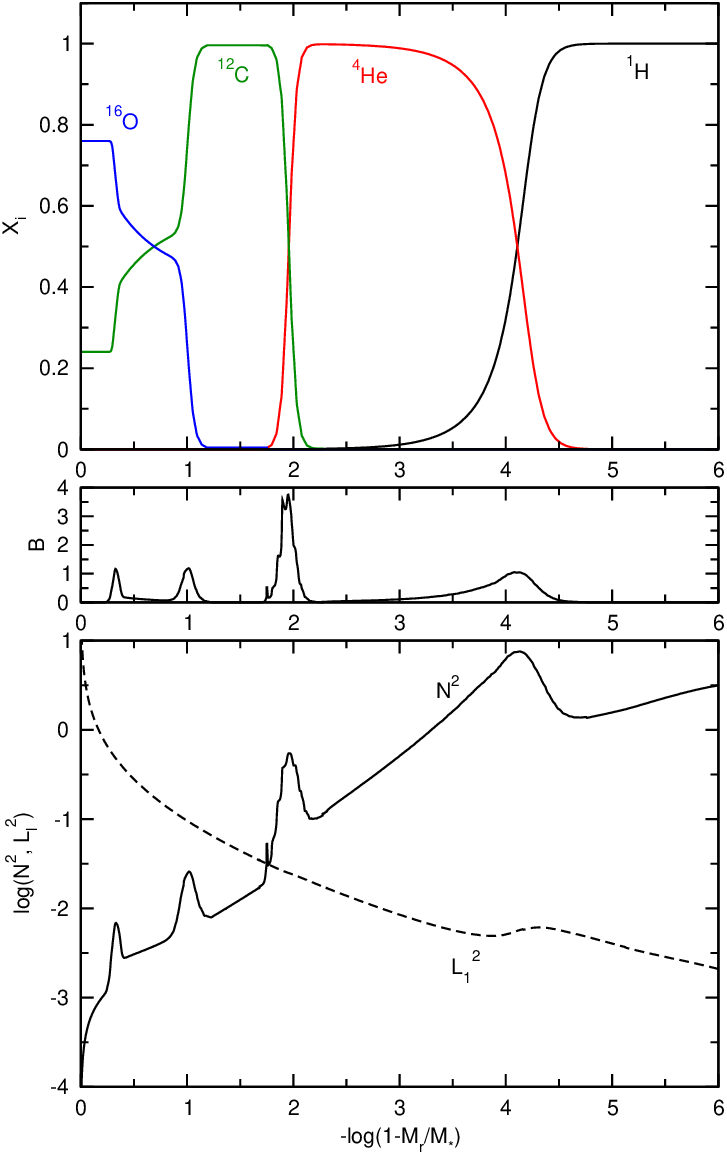}  
\caption{Same  as  in Fig.  \ref{x_b_n_LPCODE}, but  for  the  case of 
         Salaris-like core chemical profiles.}
\label{x_b_n_Salaris-like}  
\end{figure}  

In  Fig. \ref{x_b_n_ramp-like}  we  show the  situation  in which  our
template model is characterized by a ramp-like core chemical structure
of the  kind used by Bradley  (1996, 1998, 2001) and  more recently by
Bischoff-Kim (2009).   In this case  the core chemical profile  is not
the result of stellar evolution calculations, but parameterized.  This
kind of chemical profiles has been widely employed in asteroseismology
of white dwarfs because it  allows a full exploration of the parameter
space regarding  the shape of  the chemical abundance profiles  in the
core. The  parameters are the central oxygen  abundance ($X_{\rm O}$),
the coordinate at which $X_{\rm O}$ starts to drop, and the coordinate
at which $X_{\rm O}$ drops to  zero --- see, e.g., Bischoff-Kim et al.
(2008). The overall  shape of the core chemical  profiles displayed in
Fig.  \ref{x_b_n_ramp-like} is substantially  simpler than that of the
chemical    profiles   characterizing    the    model   depicted    in
Fig. \ref{x_b_n_LPCODE}.   Note the presence of  a chemical transition
region of  C and He.  This  is at variance with  the chemical profiles
produced  by  {\tt  LPCODE},  which  are  characterized  by  a  triple
transition region of oxygen, carbon,  and helium (see Fig. \ref{x_b_n_LPCODE}).  The
shape of the C/He chemical interface in this model is set by diffusion
parameters,  which  were chosen  to  match  the  pulsation periods  of
G 117$-$B15A (Bischoff-Kim et al. 2008).  It is worth noting that this
transition   region  produces   the   most  prominent   bump  in   the
Brunt-V\"ais\"al\"a frequency  (see the Ledoux  term B in  the central
panel of  Fig.  \ref{x_b_n_ramp-like}).  We mention  that the presence
of a thick pure carbon buffer like that assumed in these models is not
expected  from  stellar  evolution  calculations.  Finally,  the  He/H
chemical transition  region has been obtained  by assuming equilibrium
diffusion (Arcoragi  \& Fontaine 1980), but without  the trace element
approximation ---  see Bischoff-Kim et  al.  (2008) for  details.  The
shape of this interface is very similar to one based on time-dependent
diffusion.  It  is worth noting, however, that  the slight differences
in the thickness and steepness  of this chemical interface between the
{\tt LPCODE}  model and the  ramp-like model lead to  a non-negligible
contribution to the  differences found in the period  spacing (and the
period themselves) of the models (see later in this section).

In  Fig. \ref{x_b_n_Salaris-like} we  display the  case of  a template
model characterized  by a  Salaris-like core chemical  structure. This
kind  of core  chemical profiles  was employed  first by  C\'orsico et
al. (2001, 2002a)  and Benvenuto et al.  (2002a,b),  and more recently
in Bischoff-Kim  et al.  (2008). Actually, the  core chemical profiles
displayed  in   Fig.   \ref{x_b_n_Salaris-like}  and   those  used  in
Bischoff-Kim et  al.  (2008)  are a close  adaptation of  the original
chemical  profiles of  Salaris et  al.  (1997).   Except for  the core
region, the rest  of the chemical profiles in this  model are the same
as in the model  depicted in Fig.  \ref{x_b_n_ramp-like}.  At variance
with the template models described  before, in this case there are two
core chemical interfaces of oxygen and carbon instead of just one. This leads to
four bumps in the Brunt-V\"ais\"al\"a frequency, as can be seen in the
lower panel in Fig. \ref{x_b_n_Salaris-like}.

Figures       \ref{x_b_n_LPCODE},      \ref{x_b_n_ramp-like},      and
\ref{x_b_n_Salaris-like}  clearly   reveal  the  profound  differences
existing   between  the  chemical   profiles  and   the  run   of  the
Brunt-V\"ais\"al\"a  frequency  of the  model  generated with  LPCODE,
i.e., by considering  the full evolution of progenitor  stars, and the
two  template models that  incorporate the  most widely  used chemical
profiles in  past and current  asteroseismological studies of  ZZ Ceti
stars.   The  differences  are  particularly noteworthy  in  the  core
chemical structure.

\begin{figure} 
\centering 
\includegraphics[clip,width=230pt]{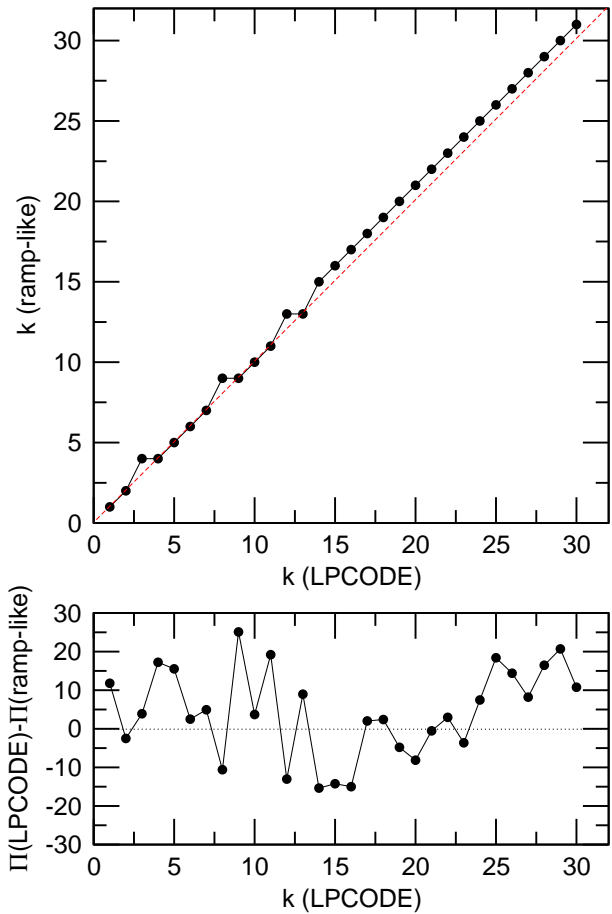} 
\caption{Comparison  between  the  template model  generated with {\tt 
         LPCODE}  (Fig.  \ref{x_b_n_LPCODE})  and  the template  model
         with    ramp-like core    chemical    profiles    (Fig. 
         \ref{x_b_n_ramp-like}). The upper panel shows the differences
         in the $k$  identification, and the lower panel  depicts  the
         differences between the matched periods.}
\label{kdiff-LPCODE-RAMP}  
\end{figure}  

\begin{figure}  
\centering  
\includegraphics[clip,width=230pt]{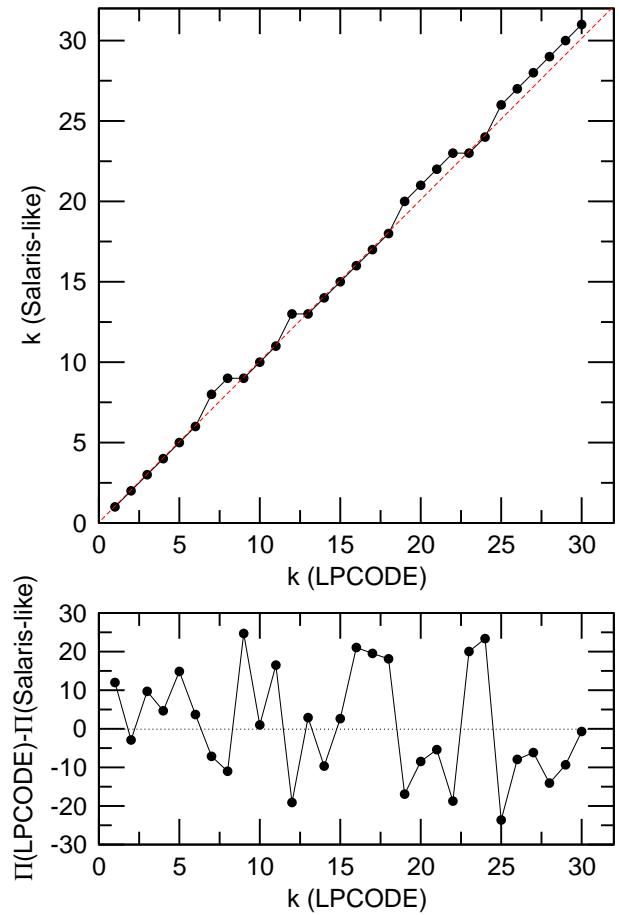}  
\caption{Same  as   in   Fig.  \ref{kdiff-LPCODE-RAMP},  but  for  the 
         comparison between  the template model  generated with LPCODE
         (Fig.  \ref{x_b_n_LPCODE})   and  the  template   model  with
         Salaris-like          core          chemical         profiles
         (Fig. \ref{x_b_n_Salaris-like}).}
\label{kdiff-LPCODE-SALARIS}  
\end{figure}  

In what follows,  we compare the pulsation properties  of our template
model with the models  having core ramp-like and Salaris-like chemical
profiles.  The  $\ell=1$ asymptotic period spacing is  largest for the
{\tt  LPCODE}  model (45.38  s),  followed  by  the Salaris-like  core
chemical profile  model (44.17  s) and by  the ramp-like  model (43.32
s).  Since these  models  have  the same  stellar  mass and  effective
temperature, the period and  asymptotic period spacing differences are
exclusively due  to the  differences in the  chemical profiles  at the
core {\sl and}  the envelope of the three  models.  In particular, the
subtle  differences  existing  in  the  shape of  the  He/H  interface
resulting  from  time-dependent   diffusion  and  that  obtained  from
equilibrium diffusion  give rise  to a non-negligible  contribution to
the difference in the asymptotic period spacing.

If   we   now   turn   the   argument   around   and   imagine   doing
asteroseismological fits  where we fix the chemical  profiles to those
found using {\tt LPCODE} and  allow the mass and effective temperature
of the  models to vary  to give us  a pulsation spectrum  that matches
that  of an  observed  pulsating white  dwarf,  the asymptotic  period
spacing  of   the  models  will  influence  the   mass  and  effective
temperature of the  best fits models. As it is  well known, the hotter
and more  massive models have smaller asymptotic  period spacings (for
modes  that are not  strongly trapped).   Since the  asymptotic period
spacing is  larger for  the LPCODE models,  the best fit  models would
have to  have larger mass and  effective temperature to  match a given
observed asymptotic  period spacing. As  a result, we would  expect to
find asteroseismological  fits that are  more massive and  hotter than
current fits.  This effect  should mainly be  observed for  rich white
dwarf pulsators, where we have  a wealth of higher $k$ (asymptotically
spaced) modes to fit.  For  G 117$-$B15A, for instance, we cannot draw
any conclusions from the asymptotic period spacings alone, as the only
3 modes  observed have  low radial overtone  and are  strongly trapped
(Bischoff-Kim et al. 2008).

The higher period spacing for the  {\tt LPCODE} model leads to a drift
to  longer and  longer periods  as we  work down  the list  of periods
toward higher  $k$ modes.  For  instance, the cumulated effect  of the
2.06 s  difference between the  {\tt LPCODE} and the  ramp-like models
results in higher $k$ periods to  differ by as much as 65 seconds (for
$k \sim 30$ if $\ell=1$ and  $k \sim 55$ if $\ell= 2$).  The practical
result  in  asteroseismological studies  of  such  a  drift to  higher
periods would  be to lead  to different $k$ identifications  of modes.
In  Figs.  \ref{kdiff-LPCODE-RAMP}  and  \ref{kdiff-LPCODE-SALARIS} we
show  how asteroseismological  fits  may be  affected.   In the  upper
panels,  we show  the differences  in $k$  identifications and  in the
bottom  panels, how  the period  of  the matched  modes differ.   Even
allowing  the $k$  identifications to  change to  find the  best match
between the  periods of  the two models  compared, we still  find that
some individual periods differ by as much as $\sim 25$ seconds.

\begin{figure*}  
\centering  
\includegraphics[clip,width=450pt]{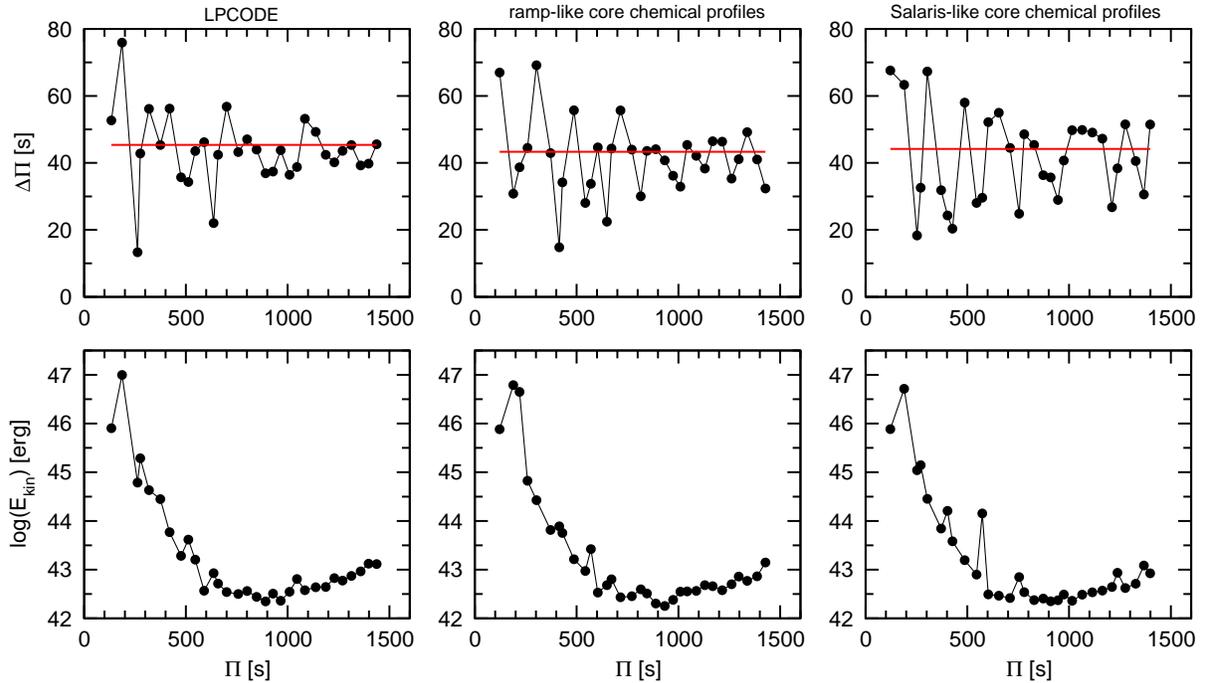}  
\caption{The  upper panels show  the forward period  spacing  and  the 
         lower panels depict the  logarithm of the oscillation kinetic
         energy  of  $\ell=  1$   modes  in  terms  of  the  pulsation
         periods. The horizontal lines in the upper panels display the
         asymptotic period spacing.}
\label{delp-ekin}  
\end{figure*}  

We  conclude our  analysis  by examining  the  forward period  spacing
($\Delta  \Pi_k\equiv  \Pi_{k+1}-   \Pi_k$)  and  the  kinetic  energy
($E_{\rm kin}$) of the three template models (Figure \ref{delp-ekin}).
The  kinetic energy  is  computed according  to  C\'orsico \&  Althaus
(2006).   The horizontal  lines in  red correspond  to  the asymptotic
period  spacing.   As  it  is  well known,  the  presence  of  density
gradients  in the  chemical transition  regions in  the interior  of a
white dwarf causes mode trapping  (Winget et al. 1981; Brassard et al.
1992a; C\'orsico et al. 2002b).  The signature of mode trapping on the
period-spacing  distribution is  the presence  of strong  minima  in a
$\Delta \Pi_k\ - \Pi_k$ diagram, in contrast to the situation in which
the star is chemically homogeneous --- see, for instance, C\'orsico \&
Benvenuto (2002).  Due to  the presence of several chemical interfaces
in the  template models,  we expect to  find clear signatures  of mode
trapping.

For  the  model  generated  with  {\tt  LPCODE},  the  $\Delta  \Pi_k$
distribution exhibits two primary minima and several secondary minima,
as can be seen in the  upper left panel of Figure \ref{delp-ekin}.  In
the case  of the  model with ramp-like  core chemical  profiles (upper
central panel)  there are also  primary and secondary minima,  but the
contrast amongst them is much less  pronounced than in the case of the
{\tt  LPCODE}  model.   Finally,  in   the  case  of  the  model  with
Salaris-like core  chemical profiles (upper  right panel) there  is no
clear  distinction between  primary  and secondary  minima of  $\Delta
\Pi_k$.  That is, all the minima are very similar.

In spite of  the complexity of the mode-trapping  pattern exhibited by
the template models,  it is possible to make  some rough inferences by
examining the values of the  kinetic energy of the modes (lower panels
of Fig.   \ref{delp-ekin}).  A close  inspection of the  plots reveals
that each  minimum of $\Delta \Pi_k$  is associated with  a maximum in
$E_{\rm kin}$ of a mode with the same radial order $k$ or differing in
$\Delta  k= \pm  1$. Modes  corresponding to  local maxima  in $E_{\rm
kin}$ are modes  partially confined to the core  regions below the O/C
and/or the C/He chemical interfaces (the O/C/He transition in the case
of the {\tt LPCODE} model),  that is, modes with amplitudes relatively
large  even  in  very deep  layers  of  the  model.  These  modes  are
potentially  useful from  an  asteroseismological point  of view.  The
remaining modes  (which have neither maxima nor  minima kinetic energy
values)  are much  less  sensitive  to the  presence  of the  chemical
transition  regions,   and  so,  they  are  of   minor  relevance  for
asteroseismolgy.

We conclude  that the  pulsation properties of  DA white  dwarf models
that incorporate our  new chemical profiles for the  core and envelope
substantially  differ from those  of models  having the  most commonly
used chemical profiles. The important  issue to be addressed now is to
asses the  impact of our new chemical  profiles on asteroseismological
period-to-period fits of real DAV stars. This step is beyond the scope
of the present work, and we defer it to future papers.

%_____________________________________________________________________

\section{Summary and conclusions}
\label{conclusions}

In  this paper  we computed  new chemical  profiles for  the  core and
envelope  of white dwarfs  appropriate for  pulsational studies  of ZZ
Ceti stars.   These profiles were  derived from the full  and complete
evolution of progenitor stars from the zero age main sequence, through
the  thermally-pulsing and  mass-loss phases  on the  asymptotic giant
branch (AGB).  These new profiles are intented for asteroseismological
studies  of ZZ  Ceti stars  that require  realistic  chemical profiles
throughout  the white dwarf  interiors. In  deriving the  new chemical
profiles,  we employed the  {\tt LPCODE}  evolutionary code,  based on
detailed  and  updated  constitutive physics.   Extra-mixing  episodes
during  central hydrogen  and helium  burning,  time-dependent element
diffusion during  the white dwarf stage  and chemical rehomogenization
of   the   inner    carbon-oxygen   composition   by   Rayleigh-Taylor
instabilities were considered. The  metallicity of progenitor stars is
$Z=0.01$.

We discussed at  some length the importance of  the initial-final mass
relationship  for  the   white  dwarf  carbon-oxygen  composition.   A
reduction  of  the  efficiency  of extra-mixing  episodes  during  the
thermally-pulsing  AGB   phase,  supported  by   different  pieces  of
theoretical and  observational evidence, yields a  gradual increase of
the hydrogen-free  core mass as evolution proceeds  during this phase.
As a  result, the  initial-final mass relationship  by the end  of the
thermally-pulsing AGB  is markedly different from  that resulting from
considering the mass of the  hydrogen free core right before the first
thermal  pulse.  We  found that  this issue  has implications  for the
carbon-oxygen composition  expected in a white  dwarf.  In particular,
the central oxygen abundance may be underestimated by about 15\% if we
assume the white  dwarf mass to be the  hydrogen-free core mass before
the  first thermal  pulse.   We  also discuss  the  importance of  the
computation of the thermally-pulsing AGB and element diffusion for the
chemical profiles expected  in the outermost layers of  ZZ Ceti stars.
In  this  sense, we  found  a strong  dependence  of  the outer  layer
chemical stratification  on the stellar mass. In  less massive models,
the intershell  region rich  in helium and  carbon --- which  is built
during the  mixing episode at  the last AGB  thermal pulse ---  is not
removed by diffusion by the time the ZZ Ceti stage is reached.

Finally, we  performed adiabatic pulsation  computations and discussed
the implications  of our new  chemical profiles for  the pulsational
properties  of ZZ  Ceti stars.   We  found that  the whole  $g-$mode
period spectrum and the  mode-trapping properties of these pulsating
white  dwarfs  as  derived   from  our  new  chemical  profiles  are
substantially  different from those  based on  the most  widely used
chemical  profiles  in   existing  asteroseismological  studies.  

We expect the best fit parameters of asteroseismological studies using
the {\tt LPCODE} chemical profiles  to differ significantly from those found
in studies made so far. Further studies will show in what way. Will we
solve  the high mass  problem with  G117$-$B15A  and Salaris-like  core
profiles (Bischoff-Kim et al. 2008) or find thicker hydrogen layers in
asteroseismological  fits,   more  in  line   with  stellar  evolution
calculations (Castanheira \& Kepler 2008)?

%_____________________________________________________________________    

\acknowledgments

Part of  this work  was supported by  AGENCIA through the  Programa de
Modernizaci\'on   Tecnol\'ogica    BID   1728/OC-AR,   by    the   PIP
112-200801-00940  grant from  CONICET, by  the AGAUR,  by  MCINN grant
AYA2008--04211--C02--01, and by the  European Union FEDER funds.  This
research has made use of NASA's Astrophysics Data System.

%_____________________________________________________________________

\end{document}